# Hybrid IP/SDN networking: open implementation and experiment management tools

Stefano Salsano[(1)], Pier Luigi Ventre[(2)], Francesco Lombardo[(1)], Giuseppe Siracusano[(1)],
Matteo Gerola[(3)], Elio Salvadori[(3)], Michele Santuari[(3)], Mauro Campanella[(2)], Luca Prete[(4)]
(1) CNIT / Univ. of Rome Tor Vergata - (2) Consortium GARR - (3) CREATE-NET – (4) ON.Lab

**Abstract** – The introduction of SDN in large-scale IP provider networks is still an open issue and different solutions have been suggested so far. In this paper we propose a hybrid approach that allows the coexistence of traditional IP routing with SDN based forwarding within the same provider domain. The solution is called OSHI – Open Source Hybrid IP/SDN networking as we have fully implemented it combining and extending Open Source software. We discuss the OSHI system architecture and the design and implementation of advanced services like Pseudo Wires and Virtual Switches. In addition, we describe a set of Open Source management tools for the emulation of the proposed solution using either the Mininet emulator or distributed physical testbeds. We refer to this suite of tools as Mantoo (Management tools). Mantoo includes an extensible web-based graphical topology designer, which provides different layered network "views" (e.g. from physical links to service relationships among nodes). The suite can validate an input topology, automatically deploy it over a Mininet emulator or a distributed SDN testbed and allows access to emulated nodes by opening consoles in the web GUI. Mantoo provides also tools to evaluate the performance of the deployed nodes.

I. INTRODUCTION

Software Defined Networking (SDN) [1] [2] is a new paradigm proposed in data networking that may drastically change the way IP networks run today. Significant use cases include Data Centers and corporate/campus scenarios. SDN applicability in wide area IP networks of large providers is being considered. At present, these networks are operated with a combination of IP and MPLS technologies. IP/MPLS control and forwarding planes are capable to operate on large-scale networks with carrier-grade quality, while SDN technology has not reached the same maturity level. The advantage of introducing SDN technology in a carrier grade IP is not related to performance improvements for current services on IP/MPLS backbones. Data Plane forwarding performances, restoration times in case of failures, several Control Plane aspects (e.g. routing convergence time) have all been optimized for the IP/MPLS backbones by the major equipment vendors in the years. We rather believe that the openness of the SDN approach simplifies the need of complex distributed Control Plane architectures and avoids proprietary implementations and interoperability issues. The new approach will facilitate the development of new services and foster innovation. The importance of Open Source in SDN is highlighted in [3] and the rising interest on white box networking [4] confirms its relevance in current and near future networking arena.

Taking the openness as the main driver for moving to SDN, the scientific and technological question "what is the best way to introduce SDN in large-scale IP Service Providers (ISP) networks?" is definitely still open and different solutions have been proposed. The OSHI (Open Source Hybrid IP/SDN) networking architecture, first introduced in [5], addresses the above question, providing an Open Source reference implementation complemented with a rich set of services and management tools.

The introduction of SDN in wide area ISP networks implies finding solutions to critical requirements and issues, such as: i) how to provide the scalability and fault tolerance required in operators' environments; ii) how to cope with the high latency in the control plane (due to the geographically distributed environment); iii) how to provide the connectivity in the Control Plane between SDN controllers and the switches in the WAN (i.e. *in-band* vs. *out-of-band* solution)

In order to support both the development/testing aspects and the evaluation of different solutions it is fundamental to have a realistic emulator platform. The platform should allow scaling up to hundreds of nodes and links, to emulate a large scale IP carrier network. Performing experiments has to be affordable for research and academic teams, not only for corporate developers. Therefore, we advocate the need of an Open Source reference node implementation and of Open Source emulation platforms. The management of these emulation platforms and the tools for setting up and controlling experiments are also non-trivial problems, which is why we propose an Open Source set of tools called Mantoo (**Man**agement **too**ls). The Mininet emulator is widely used by the SDN community, but its fidelity cannot be taken for granted especially for large scale topologies. The emulation over distributed SDN testbeds is in general more scalable and can allow to gather more realistic details on specific performance aspects. Mantoo is able to support both cases with a unified design and modelling approach.

The main contributions of this paper are:
1. The design of a hybrid IP/SDN architecture called Open Source Hybrid IP/SDN (OSHI).
2. The design and implementation of a hybrid IP/SDN node made of Open Source components.
3. Mantoo, a set of management tools to deploy and test the OSHI framework and services on Mininet emulator and on distributed SDN testbeds
4. Evaluation of some performance aspects of the OSHI prototype implementation over distributed SDN testbeds.

On top of the proposed OSHI framework and Mantoo tools the researcher/developer is able to design and deploy new services and to experiment on SDN Control Plane solutions with a minimal effort. The paper is structured as follows: section II describes the scenarios related to the introduction of SDN in IP Service Provider networks; section III defines the main concepts of the proposed hybrid IP/SDN networking architecture; section IV provides a detailed description of the

OSHI nodes implementation and of the services that such a solution can offer; section V identifies some limitations of current SDN ecosystem along with the needed extensions, it also reports how our framework is being used to experiment on new services; section VI describes the Mantoo suite, that allows to design, deploy and control experimental topologies in a local emulator (Mininet) or on distributed testbeds, supporting the collection of performance measurements; section VII provides an evaluation of some performance aspects; section VIII reports on related work and explains the main differences with respect to our previous work; in section IX we draw some conclusions and highlight how we are porting OSHI over white box switches, potentially stepping from experiments to production networks.

The source code of all the components of the OSHI node prototypes and of the Mantoo suite is freely available at [6]. To facilitate the initial environment setup, the whole OSHI and Mantoo environments have been packaged in a ready-to-go virtual machine, with pre-designed example topologies up to 60 nodes. To the best of our knowledge, there is no such hybrid IP/SDN node available as Open Source software, nor an emulation platform with a set of management tools as rich as the Mantoo suite.

## II. SDN Applicability in IP Providers Networks

SDN is based on the separation of the network Control Plane from the Data Plane. An external SDN controller can (dynamically) inject rules in SDN capable nodes. According to these rules the SDN nodes perform packet inspection, manipulation and forwarding, operating on packet headers at different layers of the protocol stack.

We focus on SDN applicability in IP Service Provider networks. Figure 1 shows a reference scenario, with a single IP provider interconnected with other providers using the BGP routing protocol. Within the provider network, an intra-domain routing protocol like OSPF is used. The provider offers Internet access to its customers, as well as other transport services (e.g. layer 2 connectivity services or in general VPNs - Virtual Private Networks). Using the terminology borrowed by IP/MPLS networks, the provider network includes a set of Core Routers (CR) and Provider Edge (PE) routers, interconnected either by point-to-point links (Packet Over Sonet, Gigabit Ethernet, 10GBE…) or by legacy switched LANs (and VLANs). The Customer Edge (CE) router is the node in the customer network connected to the provider network. Most often, an ISP integrates the IP and MPLS technologies in its backbone. MPLS creates *tunnels* (LSP – Label Switched Path) among routers. On one hand, this can be used to improve the forwarding of regular IP traffic providing: i) traffic engineering, ii) fault protection iii) no need to distribute the full BGP routing table to intra-domain transit routers. On the other hand, MPLS tunnels are used to offer VPNs and layer 2 connectivity services to customers. In any case, the commercial MPLS implementations are based on traditional (vendor-locked) control plane architectures that do not leave space for introducing innovation in an open manner. As a matter of fact, in case of complex services involving the MPLS control plane, IP Service Providers rely on single-vendor solutions. The management of large-scale IP/MPLS network is typically based on proprietary (and expensive) management tools, which, again, constitute a barrier to the innovation.

Let us consider the migration of an IP/MPLS based Service Provider network to SDN. CR and PE routers could be replaced by SDN capable switches, on top of which the provider can realize advanced and innovative services. The migration paths should foresee the coexistence of IP and SDN based services, resembling the current coexistence of IP and MPLS. We define as *hybrid IP/SDN* a node that can operate both at IP level by keeping a traditional distributed routing intelligence and at SDN level, under the instructions of a SDN controller. This is opposed to a *pure SDN* node in which all routing logic is ran outside the node in the SDN controller. A hybrid IP/SDN network is composed of hybrid IP/SDN nodes, as well as by traditional IP routers and legacy layer 2 switches. According to the taxonomy defined in [7], this approach can be classified as "Service-Based" or "Class-Based" Hybrid SDN (depending on how the IP and SDN based services are combined). In this scenario the hybrid IP/SDN nodes are capable of acting as plain IP routers (running the legacy IP routing protocols), as well as SDN capable nodes, under the control of SDN controllers.

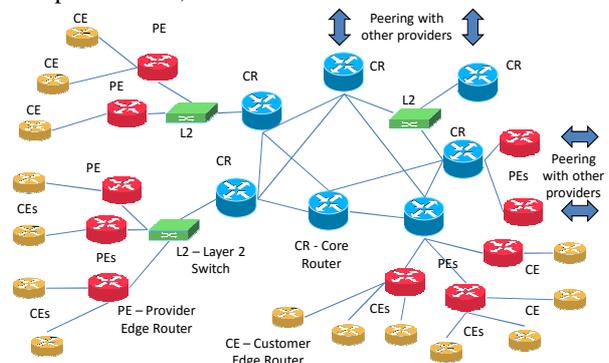

Figure 1. Reference scenario: an IP provider network

## III. Proposed Hybrid IP/SDN Architecture

In the IP/MPLS architecture there is a clear notion of the MPLS tunnels, called Label Switched Paths (LSPs). In a SDN network several types of tunnels or, more generically, *network paths* can be created, leveraging on the ability of SDN capable nodes to classify traffic based on various fields such as MAC or IP addresses, VLAN tags and MPLS labels. Since there is no standard established terminology for such concept, we will refer to these paths as *SDN Based Paths* (SBP). A SBP is a *virtual circuit* which is setup using SDN technology to forward a specific *packet flow* between two end-points across a set of SDN capable nodes. The notion of packet flow is very broad and it can range from a *micro-flow* i.e. a specific TCP connection between two hosts, to a *macro-flow* e.g. all the traffic directed towards a given IP subnet. As highlighted before, a flow can be classified looking at the headers at different protocol levels.

We address the definition of the hybrid IP/SDN network by considering: i) mechanisms for the coexistence of regular IP traffic and SBPs; ii) the set of services that can be offered using the SBPs; iii) ingress traffic classification mechanisms.

Let us consider the coexistence of regular IP traffic and SDN based paths on the links among hybrid IP/SDN nodes. A SDN approach offers a great flexibility, enabling the classification of the packets through a "cross-layer" approach, by considering packet headers at different protocol levels (MPLS, VLANs, Q-in-Q, Mac-in-Mac and so on). Therefore, it is possible to specify a set of conditions to differentiate the packets to be delivered to the IP forwarding engine from the ones that belong to SBPs. In general, these conditions can refer to different protocol headers and can be in the form of whitelists or blacklists, changing dynamically, interface by interface. This flexibility may turn into high complexity and the risk of misconfigurations and routing errors should be properly taken into account (see [8]). Without preventing the possibility to operate additional mechanisms for the coexistence of IP and SDN services in a hybrid IP/SDN network, we propose MPLS tagging as the preferred choice and have used it in our prototype implementation. In fact, using MPLS as forwarding plane technology is known to be scalable up to carrier-grade WANs. We have also considered simple VLAN tagging as a sub-optimal choice and have used it in a simpler prototype (see [5][9]). Simple VLAN tagging limits the number of SBPs on a link to 4096. Moreover, if legacy VLAN services needs to be supported on the links among the OSHI nodes, the VLAN label space needs to be partitioned, reducing the maximum number of SBPs and complicating the service management process.

In a SDN solution for wide area networks there is the problem to setup the connectivity between SDN controllers and OF capable switches. This is usually solved with out-of-band communication channels, as it is complicated to reliably "bootstrap" and maintain the connectivity using the data plane links with a centralized control. A key advantage of the coexistence approach in the proposed OSHI architecture is the possibility to use traditional IP routing and forwarding for the Control Plane connectivity between SDN controllers and OF Capable switches. This approach avoids the needs of out-of-band communication channels for the Control Plane.

Let us now consider the services and the features that can be offered by a hybrid IP/SDN network. As primary requirements we assume three main services/functionalities: (i) virtual private networks (Layer 2 and Layer 3), (ii) traffic engineering, (iii) fast restoration mechanisms. Moreover, the architecture should facilitate the realization of new services and the development of new forwarding paradigms (for example Segment Routing [22]) without the need of introducing complex and proprietary control planes.

As for the traffic classification, the ingress PEs need to classify incoming packets and decide if they need to be forwarded using regular IP routing or if they belong to the SBPs. The egress edge router extracts the traffic from the SBPs and forwards it to the appropriate destination. We considered (and implemented in our platform) two approaches for the ingress classification: i) classification based on physical access ports; ii) classification based on VLAN tags. Other traffic classifications, e.g. based on MAC or IP source/destination addresses can be easily implemented without changing the other components.

IV. DETAILED DESIGN OF THE HYBRID IP/SDN SOLUTION

In this section we present the detailed design and the implementation of the proposed architecture. We describe the Open Source tools that we have integrated and how their practical limitations have been taken into account to deliver a working prototype. We first introduce the high level architecture of an OSHI node (IV.A) and the basic services we provide (IP Virtual Leased Line and Pseudo-wires, IV.B). Then we describe the use of MPLS labels to realize SDN Based Paths (SBPs) and to support the coexistence between IP based forwarding and SBP forwarding. We show the design challenges of the MPLS based implementation, partly due to the inherent limitations of the current OpenFlow standards, partly to the shortcomings of the Open Source tools that we have integrated.

A. *OSHI High Level Node Architecture*

The proposed OSHI node combines an OpenFlow Capable Switch (OFCS), an IP forwarding engine and an IP routing daemon. The OFCS component is implemented using Open vSwitch (OVS) [31], the IP forwarding engine is the Linux kernel IP networking and Quagga [16] acts as the routing daemon. The OpenFlow Capable Switch is connected to the set of physical network interfaces belonging to the integrated IP/SDN network, while the IP forwarding engine is connected to a set of virtual ports of the OFCS, as shown in Figure 2.

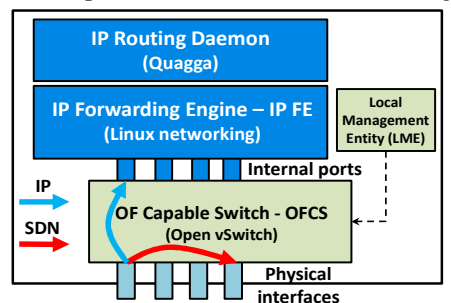

Figure 2. OSHI Hybrid IP/SDN node architecture

The virtual ports that interconnect the OFCS with the IP forwarding engine are realized using the *Internal Port* feature offered by Open vSwitch. Each internal port is connected to a physical port of the IP/SDN network, so that the IP routing engine can reason in term of the virtual ports, ignoring the physical ones. The OFCS differentiates among regular IP packets and packets belonging to SDN Based Paths. By default, it forwards the regular IP packets from the physical ports to the internal ports, so that they can be processed by the IP forwarding engine, controlled by the IP routing daemon. This approach avoids the need of translating the IP routing table into SDN rules to be pushed in the OFCS table, at the price of a small performance degradation for the packets that needs to be forwarded at IP level. In fact, these packets cross the OFCS switch twice. It is possible to extend our implementation to consider the mirroring of the IP routing table into the OFCS table. Mapping a static snapshot of the IP routing table into a set of SDN rules in the OFCS is relatively easy (the rewriting of source and destination MAC addresses needs to be included in the rules and the MAC addresses of the next hops needs to be discovered beforehand). The difficult challenge is to take into account the dynamic aspects, as the rules should be updated in a timely way following route additions, updates, deletions. Therefore in the OSHI prototype presented in this work this feature is left out for future work In [17] we described a prototype solution that mirrors the routes installed by OLSR in real time (for a specific set of IP destinations), mapping them in OpenFlow rules.

An initial configuration of the OFCS tables is needed to connect the physical interfaces and the internal interfaces, in order to support the OFCS-to-SDN-controller communication and some specific SDN procedures (for example to perform layer 2 topology discovery in the SDN controller). A Local Management Entity (LME) in the OSHI node takes care of these tasks. In our setup, it is possible to use an "in-band" approach for the OFCS-to-SDN-controller communication, i.e. using the regular IP routing/forwarding and avoiding the need of a separate out-of-band network. Further details and the block diagram of the control plane architecture of OSHI nodes are reported in [9].

### B. *OSHI basic services: IP VLL and L2 PW*

We designed and implemented two basic services to be offered by OSHI networks: the "IP Virtual Leased Line" (IP VLL) and the Layer 2 "Pseudo-wire" (L2 PW or PW in short) see Figure 3. They belong to the class of Virtual Leased Line services [28], which are a fundamental part of the offering of large-scale IP Service Providers. VLL services can be used to carry bandwidth guaranteed applications (e.g. real time communications) or to support VPN solution (e.g. interconnect different sites of a company through the ISP WAN). Both services are offered between end-points in Provider Edge routers, the end-points can be a physical or logical port (i.e. a VLAN on a physical port) of the PE router connected to a Customer Edge (CE). The interconnection is realized in the core hybrid IP/SDN network with an SBP using MPLS labels.

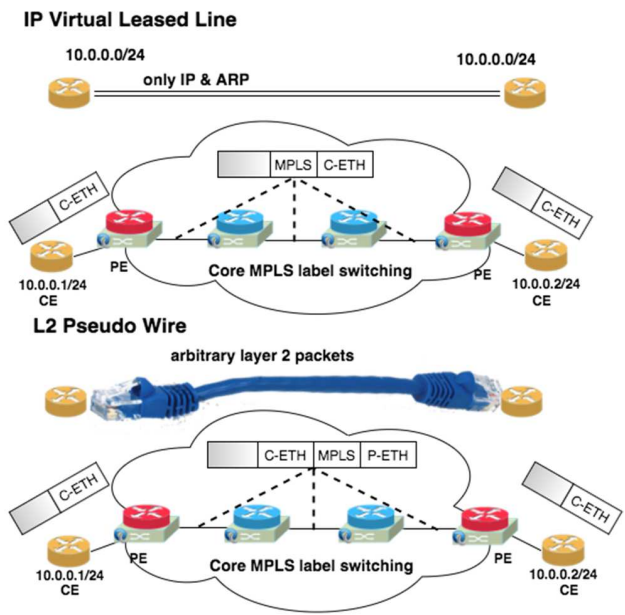

Figure 3. IP VLL and L2 PW services

The proposed IP VLL service guarantees to the IP end-points to be directly interconnected as if they were in the same Ethernet LAN and sending each other IP and ARP packets. It is not meant to allow the served SBP end-points to send packets with arbitrary Ethertype (e.g. including VLAN packets). The original source and destination MAC addresses, shown as "C-ETH" (C stands for Customer) in the headers of the packets in Figure 3, are preserved in the transit along the network core. This may cause problems if legacy L2 switches are used to interconnect OSHI nodes, therefore our implementation of IP VLL service can only work if all edge and core nodes are OSHI capable and are directly connected to each other, without legacy intermediate switches in between. As a solution to interwork with legacy switches, one could implement MAC address rewriting replacing the customer addresses with the addresses of the ingress and egress PEs or on a hop-by-hop case. This is rather complex to realize and to manage, because the egress node should restore the original MAC addresses (using the tag as key). There is the need to exchange and then maintain additional state information per each SBP in the egress nodes, so we did not implement this solution. In our prototype and experiments, if legacy switches are present in the network, the L2 PW service rather than the IP VLL service should be used.

The L2 PW service is also known as "Pseudowire Emulation Edge to Edge" (PWE3), described in RFC 3985 [24]. It provides a fully transparent cable replacement service: the endpoints can send packets with an arbitrary Ethertype (e.g. including VLAN, Q-in-Q). As shown in Figure 3, the customer Ethernet packet is tunneled into a new Ethernet packet (whose header is indicated as P-ETH) and then a MPLS header is added. This approach solves the interworking issues with legacy L2 networks related to customer MAC addresses exposure in the core.

### C. *OSHI - MPLS based approach*

In this subsection we illustrate the detailed aspects of the proposed solution based on MPLS. The use of MPLS labels enables the establishment of up to $2^{20}$ (more than $10^6$) SBPs on each link, providing the required scalability. The MPLS label space can be partitioned in order to have an ordered coexistence with other MPLS based services in the provider WAN. We describe the implementation of IP VLL and PW services, in both cases the MPLS solution does not interfere with VLANs that can potentially be used in the links between OSHI nodes.

*1) Coexistence mechanisms*

The coexistence of regular IP service (best effort traffic) and SDN services (using SDN Based Paths) is assured using the Ethertype field of the L2 protocol. This corresponds to one of the mechanisms that can be used in the IP/MPLS model: regular IP traffic is carried with IP Ethertype (0x0800), while SBPs are carried with MPLS Ethertypes (0x8847 and 0x8848). Using OpenFlow multi-table functionality, our solution supports the coexistence of IP and MPLS traffic types, as shown in Figure 4. Table 0 is used for regular IP, ARP, LLDP, BLDP, etc., table 1 for the SBPs. In particular, Table 0 contains: i) a rule that forwards the traffic with Ethertype 0x8847 (MPLS) to Table 1; ii) only for IP VLL a rule that forwards the traffic with Ethertype 0x8848 (Multicast MPLS) to Table 1; iii) the set of rules that "bridge" the physical interfaces with the internal ports and vice versa; iv) two rules that forward the LLDP and BLDP traffic to the controller. Table 1 contains the set of rules that forward the packets of the SBPs according to the associated IP VLL or PW service. The coexistence in Table 0 is assured through different levels of priority. The IP VLL service needs both the rules associated to unicast and multicast MPLS Ethertype (more details below), while the PW service only needs a rule matching the unicast MPLS Ethertype.

We consider two MPLS based tunneling mechanisms: plain IP over MPLS ([23], here referred to as IPoMPLS) and Ethernet over MPLS (EoMPLS [24] [25]). The IPoMPLS tunneling is used for the IP VLL service. The EoMPLS tunneling can support the relaying of arbitrary layer 2 packets, providing the L2 PW service [24].

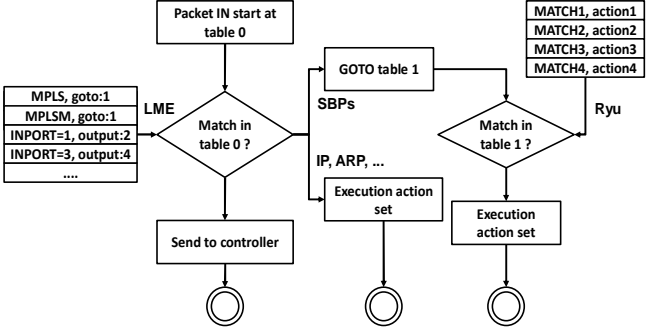

Figure 4. Packet processing in the OFCS flow tables

*2) Ingress classification and encapsulation mechanisms*

As for the ingress classification functionality in a PE router, it can be either based on the physical input port or on the incoming VLAN tag. We use the input port to classify untagged traffic as *regular IP* traffic or as belonging to a SBP end-point (of an IP VLL or PW). For the VLAN tagged traffic entering in a physical port of a PE router, each VLAN tag can be individually mapped to a SBP end point or assigned to regular IP traffic. For the untagged traffic, the implementation of the ingress classification is realized within the OFCS of the OSHI Provider Edge nodes. In fact, by configuring rules in the OFCS, it is possible to map the untagged traffic on an ingress physical port to an internal port (for regular IP) or to a SBP. For the tagged traffic, the incoming classification relies on the VLAN handling of the Linux networking: each VLAN tag x can be mapped to a virtual interface eth0.x that will simply appear as an additional physical port of the OFCS.

Let us analyze the encapsulation mechanisms. The left half of Figure 5 shows the encapsulation realized by the OSHI-PE node for the IP VLL service. C stands for Customer, the ingress direction is from customer to core, egress refers to the opposite direction. This solution follows the IPoMPLS approach, in which a MPLS label is pushed within an existing frame. In this case an input Ethernet frame carrying either an IP or an ARP packet, keeps its original Ethernet header, shown as C-ETH in Figure 5. As we have already discussed, this solution has the problem of exposing the customer source and destination MAC addresses in the core. Moreover, note that the MPLS Ethertype (0x8847) overwrites the existing Ethertype of the customer packets. This does not allow the distinction between IP and ARP packets at the egress node. A solution would be to setup two different bidirectional SBPs: one for the IP and one for the ARP packets. In order to save label space and simplify the operation we preferred to carry IP packets with the MPLS Ethertype and to (ab)use multicast MPLS Ethertype (0x8848) to carry the ARP packets. With this approach, the same MPLS label can be reused for the two SBPs transporting IP and ARP packets between the same end-points.

The "Ethernet over MPLS" (EoMPLS) encapsulation [25] represents the most efficient approach to implement the PW service. As shown in the right side of Figure 5, EoMPLS encapsulates the customer packet including its original Ethernet header in an MPLS packet to be carried in a newly generated Ethernet header. Unfortunately, we require a solution that can be implemented using an Open Source switch and we would like to have a solution that can be fully controlled by OpenFlow. The OpenFlow protocol and most OpenFlow capable switches (including Open vSwitch that we are using for our prototype) do not natively support EoMPLS encapsulation and de-capsulation. A similar issue has been identified in [36], in which the authors propose to push an Ethernet header using a so called "input Packet Processing" (iPProc) function before handing the packet to a logical OpenFlow capable switch that - in turn - will push the MPLS label. Obviously this requires a switch with an "input Packet Processing" function capable of pushing an Ethernet header into an existing Ethernet packet. Note that this process is not fully controlled with the OpenFlow protocol, as OpenFlow does not support the pushing of an Ethernet header. We cannot directly follow this approach, as Open vSwitch is not capable of pushing Ethernet headers. The right half of Figure 5 shows the approach that we have followed, relying on GRE encapsulation. P stands for Provider and it indicates the headers added/removed by the PE. A packet in the PE is processed in four steps (shown as i1 to i4 in the ingress direction from the CE towards the core and as e1 to e4 in the egress direction from the core toward a customer. The GRE encapsulation introduces an additional overhead (20 bytes for P-IP and 4 bytes for GRE headers) to the standard EoMPLS, but it allowed us to rely on Open Source off-the-shelf components.

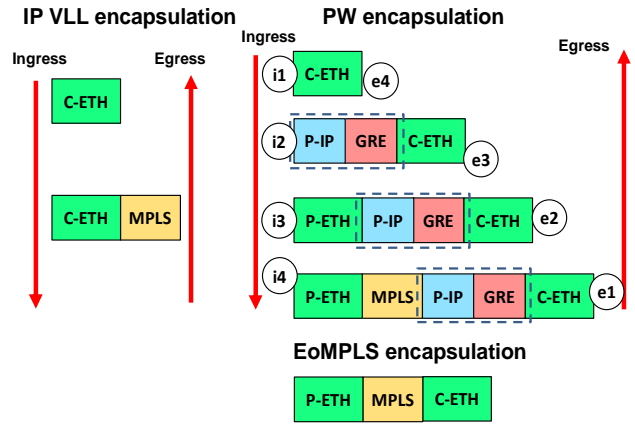

Figure 5. IP VLL and L2 PW tunneling operations at the Provider Edges. The EoMPLS encapsulation format is shown as a reference.

The implementation of the proposed approach required a careful design, whose result is shown in Figure 6. A new entity called ACcess Encapsulator (ACE) is introduced in order to deal with the GRE tunnel at the edges of the pseudo wire tunnel. The detailed design is further analyzed in subsection IV.D.

With this approach it is possible to rewrite the outer source and destination MAC addresses in the core OSHI network, so that they can match the actual addresses of the source and destination interfaces on the OSHI IP/SDN routers. This allows the support of legacy Ethernet switched networks among the OSHI IP/SDN routers, which can be an important requirement for a smooth migration from existing networks.

Both the IP VLL and PW services are realized with SBPs that switch MPLS labels between two end-points (in both directions). We used the Ryu [38] controller, the SBPs are setup using a python script called VLLPusher. The script uses the Ryu Topology REST API of to retrieve the shortest path that interconnects the SBP end-points. It allocates the MPLS labels and then uses the Ofctl REST API to setup the rules for packet forwarding and MPLS label switching. In the setup of a PW service the MAC rewriting actions are added, using the addresses of the OSHI nodes as the outer MAC addresses.

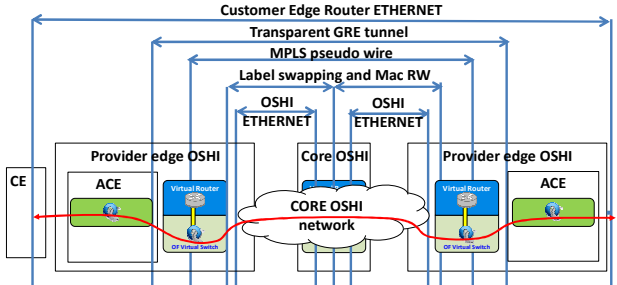

Figure 6. PW implementation in the OSHI node prototype

*3) Requirements on protocol and tools versions*

The MPLS solution needs at least OpenFlow v1.1, which makes possible to handle MPLS. Both the SDN controller and the SDN Capable Switch need to support at least OF v1.1 (most controller and switches jumped from OF v1.0 to v1.3). Considering our tools, an Open vSwitch version compliant with OF v1.3 has been released in summer 2014, making it possible to start the implementation of the MPLS based approach.

*4) The Virtual Switch Service (VSS)*

The PW service can be used as a building block for creating more complex services, like for example the Virtual Switch Service (VSS). While a PW service instance bridges two layer 2 end-points, the VSS service bridges a set of end-points into a virtual layer2 switch. The ports of a VSS instance correspond to an arbitrary set of ports of the Provider Edge nodes. This service is called Virtual Private LAN Service (VPLS) in RFC 4761 [37]. A VSS provides the same VPLS service described in the RFC but its implementation is based on SDN and does not exploit other control plane functionalities, therefore we renamed it.

The VSS is based on the L2 PW service, because the IP VLL service does not provide a transparent forwarding of layer 2 packets. To implement the VSS service, a set of PWs connect the end-points to *branching points* in the OSHI network. A virtual layer 2 switch instance, called Virtual Bridging Point (VBP), is allocated in the branching points to bridge the packets coming from the PWs.

A VSS instance is deployed in three steps: i) branching point selection; ii) VBP deployment; iii) VBP interconnection. In the first step, a python script called VSSelector retrieves the topology from the controller and then chooses the branching points, i.e. the OSHI nodes that will host the VBPs. In the second step according to the output of VSSelector the VBP are deployed as additional instances of Open vSwitch in the selected OSHI nodes (see subsection IV.D for implementation details). The final step is the deployment of the PWs that will interconnect the CEs to the VBPs and the VBPs among each other. We provide two versions of the branching point selection (first step above): i) un-optimized; ii) optimized. In the un-optimized version a single node is randomly selected in the topology and used to deploy the virtual bridge. For the optimized version, finding the optimal topology to implement a VSS corresponds to the minimal Steiner tree problem [39]. We implement the heuristic defined in [40] to find an approximate solution. Then, using the tree topology obtained from the heuristic, a VBP is deployed in each branching point of the tree. In both the un-optimized and optimized version, the VBPs are connected each other and with end-points with direct Pseudo Wires. In this way the packets enters the VBPs only in the branching points.

D. *OSHI detailed node architecture*

In order to support the PW and VSS services, the architecture of an OSHI node needs to be more complex with respect to the high level architecture shown in Figure 2. Figure 7 provides a representation of the proposed solution for the PE nodes. As discussed above, the difficult part is the support of encapsulation and de-capsulation in the OSHI PE nodes, for which we resorted to use GRE tunnels (see the right side of Figure 5). The different encapsulation steps in the ingress (i1-i4) and egress direction (e1-e4) are represented using the same numbering of Figure 5. The OF Capable Switch only handles the push/pop of MPLS labels, while the ACE handles the GRE encapsulation. The ACE is implemented with a separate instance of Open vSwitch, in particular we have an ACE instance running in a separate Linux network namespace [34] for each customer. For each PW, the ACE has two ports: a "local" port facing toward the CE locally connected to the PE node and a "remote" one facing towards the remote side of the PW. The remote port is a GRE port provided by OVS, therefore the ACE receives the customer layer 2 packets on the local ports and sends GRE tunneled packets on the remote port (and vice-versa). The interconnection of OFCS ports and ACE ports (the endpoints of the yellow pipes in Figure 7) are realized using the concept of Virtual Ethernet Pair [34] offered by the Linux Kernel.

Differently from the internal ports (shown on the right side of Figure 7), the Virtual Ethernets are always associated in pairs. In our case, for each PW two Virtual Ethernet pairs are needed, one pair is used to connect the CE port of OFCS with the local port of ACE, another pair to connect the remote port of the ACE with the physical ports towards the remote side of the PW. Three virtual Ethernet endpoints are used as plain switch ports (two belong to the OFCS, one to the ACE), the last one, on the ACE, is configured with an IP address and it is used as the endpoint of the GRE tunnel (Virtual Tunnel Endpoint, i.e. VTEP). These IP addresses are not globally visible, but they have a local scope within the network namespaces associated to the customer within all the OSHI nodes. This approach greatly simplifies the management of the services, as the same addresses for the GRE VTEP can be reused for different customers. As a further simplification, static ARP entries are added on the Virtual Ethernet for each remote tunnel end (remote VTEP). For each customer, a simple centralized database of IP and MAC addresses (used for GRE tunnels) is needed.

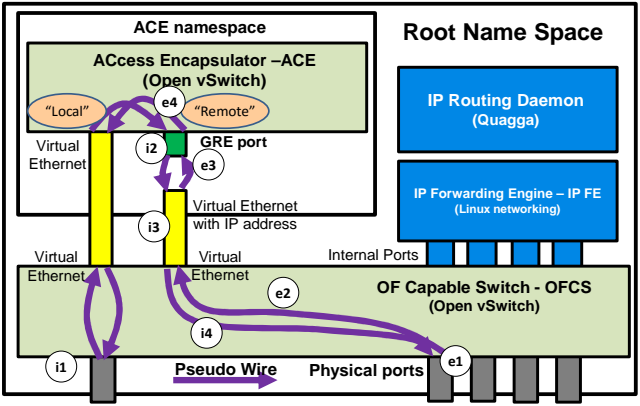

Figure 7. OSHI-PE architectural details

Proper OpenFlow rules needs to be setup in the OF Capable Switch to ensure the transit of packets. On the access port (i1) these rules are provided by the LME at the time of the ACE creation, while in the i4 and e2 cases they are pushed by the OpenFlow Controller during the PW establishment.

As discussed above, an instance of ACE in the PE node is used to handle all the PWs of a single customer and runs in a private network namespace. In addition we had to configure a private folders tree for each ACE instance, as it is needed to guarantee proper interworking of difference instances of OVS in the same PE node.

Coming to the implementation of the VSS, the internal design of an OSHI node that hosts a VSS Bridging Point (VBP) is shown in Figure 8. The design is quite similar to the one analyzed before for the PW encapsulation. A VBP is implemented with an OVS instance that does not have local ports, but only remote ones. A VPB instance represents a bridging point for a single VSS instance and it cannot be shared among VSS instances.

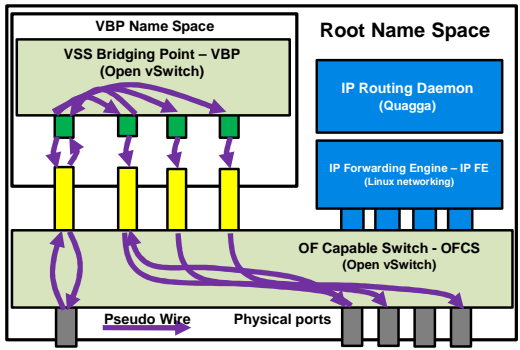

Figure 8. An OSHI node that hosts a bridging point for a VSS

*1) Considerations on alternative design choices*

Considering that a single instance of Open vSwitch can support several independent switches, a simpler design would consists in implementing the ACEs shown in Figure 7 as separate switches within the same Open vSwitch instance that runs the OFCS. For N customers, this solution would use one OVS instance instead of N and only the root network namespace instead of N additional namespaces, reducing the memory requirements versus the number of customers. The drawback of this solution is that handling the GRE tunnels of all customers in the same network namespace requires the management of disjoint IP numbering spaces for the tunnel endpoints of different customers. In addition, the separate namespaces allow to turn the ACE in a "Virtual Router" by including an instance of a routing daemon (Quagga) in its network namespace. Such a virtual router is the basic component of Layer 3 VPN services that could complement the Layer 2 PW and VSS services realized so far. With the choice of the more complex design we tradeoff scalability with simplification of the service management and easier development of new services.

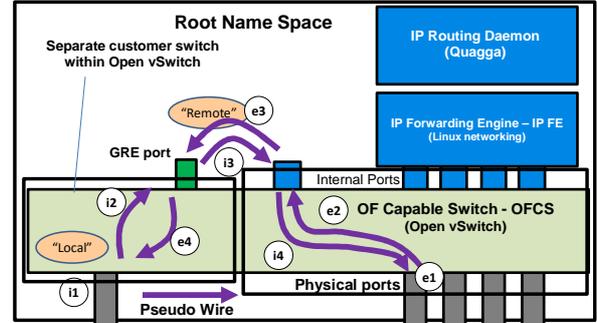

Figure 9. PW implementation design without ACE

A second consideration is that the handling of GRE tunneling has been recently introduced in Linux kernels. This can lead to a simpler design for tunneling that does not require the ACE nor the use of the GRE module provided by Open vSwitch, as shown in Figure 9. Anyway, this solution has the same drawbacks discussed above in terms of management of IP addresses for the tunnel endpoints, because there are not separate network namespaces for the customers, and cannot be easily extended to support Layer 3 services.

V. OSHI: GAP ANALYSIS, ONGOING AND FUTURE WORK

The solution for PW encapsulation described in section IV.D is based on GRE tunneling performed by the ACE. It has been designed as a replacement of the more efficient Ethernet over MPLS (EoMPLS) encapsulation specified in [24], which cannot be realized by the current version of Open vSwitch. The GRE tunneling introduces a transport and a processing overhead. The former is 20 (IP header) + 16 (GRE header) bytes for each packet, while the latter depends on the implementation architecture. Our solution (shown in Figure 7) is not meant to be highly efficient but only to demonstrate the feasibility of the approach with a working component. We do not plan to improve the efficiency of the solution, rather we believe that native Ethernet over MPLS (EoMPLS) encapsulation should be provided by open source switches and we are considering to extend the Open vSwitch to support EoMPLS.

Assuming that a switch supports EoMPLS, a second important gap to be filled is the lack of support for such tunneling operations in the OpenFlow protocol. Note that the lack of encapsulation support in OpenFlow does not only concern EoMPLS, but also other tunneling solutions like GRE, VXLAN. The only tunneling solution currently supported by OpenFlow is the PBB (Provider Backbone Bridges, also known as "mac-in-mac"), but this solution is not supported by Open vSwitch. For GRE and VXLAN, using OpenFlow it is possible to control packets already tunneled (and specific matches have been introduced in OF 1.4 for VXLAN), but it is not possible to control the encapsulation (i.e. pushing the GRE, VXLAN headers) and de-capsulation (i.e. popping the header) operations. Currently, external tools are needed to manage the GRE or VXLAN tunnel end-points (e.g. using the switch CLIs - Command Line Interfaces or switch specific protocols, like ovsdb-conf for Open vSwitch), with added complexity in the development, debug and operations. Extending OpenFlow protocol with the capability to configure the tunneling end-points would be a great simplification in the management of SDN based services.

The OSHI solution is an open starting point to design and implement additional "core" functionality and user oriented services. As for the core functionality we are considering traffic engineering mechanisms and implemented a flow assignment heuristic for optimal mapping of PWs with required capacity on the core OSHI links. As for additional services, we are considering Layer 3 VPNs based on the PW service. Following the same approach used for the VSS service, the idea is to deploy virtual router instances within the OSHI nodes that can exchange routing information with routers in the CE nodes. Finally, we are working on an Open Source implementation of Segment Routing [22] on top of OSHI [26]. This last scenario is a good example of how the proposed framework facilitates the implementation of new services and forwarding paradigms. All these ongoing efforts are reported on the OSHI web page [6], with links to documentation and source code.

## VI. Mantoo: Management Tools for SDN/NFV experiments on Mininet and Distributed SDN testbeds

Mantoo is a set of Open Source tools meant to support SDN experiments both over Mininet and over distributed testbeds. Mantoo is able to drive and help the experimenters in the different phases that compose an experiment: design, deployment, control and measurement, as described in the next subsections. Mantoo includes: a web based GUI called Topology3D (Topology and Services Design, Deploy and Direct, Figure 10), a set of scripts to configure and control emulators or distributed testbeds; a set of scripts for performance measurements. The overall Mantoo workflow is represented in Figure 11. Using the Topology3D, the user can design its experiment in terms of physical topology and services, start the deployment of the topology and run the experiments exploiting the provided measurement tools. The design of Mantoo and of its components is modular and it can be easily extended to support scenarios that go beyond the use cases of our interest.

### A. Design Phase

The Topology3D offers a web GUI to design a network topology and to configure the services for an experiment (see Figure 10). It consists of a JavaScript client and a Python back-end. A link to a public instance of the Topology 3D can be accessed from [6]. The Topology3D is meant to be an extensible framework that can support different models of topology and services. A model corresponds to a technological domain to be emulated and is characterized by the set of allowed node types (e.g. routers, switches, end-hosts), link types, service relationships and related constraints.

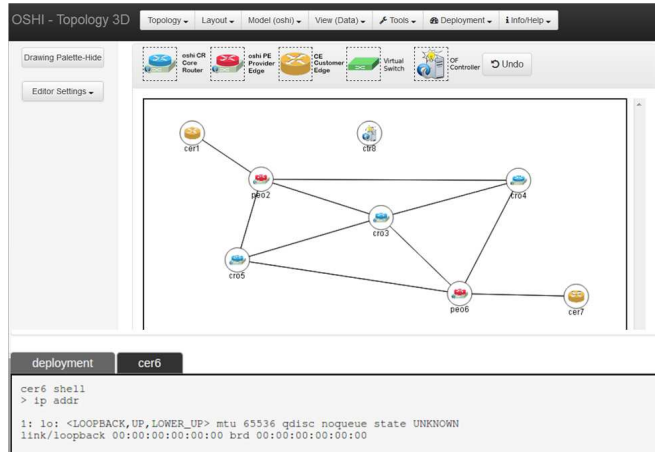

Figure 10. The Topology3D (Topology and Services Design, Deploy & Direct) web Graphical User Interface

As shown in Figure 11 the input to Topology3D is a textual description of the model. The model description is used to configure the topology designer page, to enforce the constraints when the user is building the topology and/or during the validation of the topology. So far, we have provided two models: 1) the OSHI topology domain, including OSHI CR and PE, , Customer Edge routers which are also used as traffic source/sinks and SDN controllers; 2) a generic layer 2 network with OpenFlow capable switches, end-nodes and SDN controllers. Each model is decomposed in a set of views. A view is a perspective of a model, which focuses on some aspects hiding unnecessary details. For example, the OSHI model is decomposed in 5 views: data plane, control plane and 3 views for the 3 services (IP VLLs, Pseudo Wires and Virtual Switches). In the data plane view, the user designs the physical topology in terms of nodes (OSHI CR and PE, Controllers, and CEs) and links; in the control plane view the user associates OSHI nodes with controllers; in the service views the user selects the end points of the services.

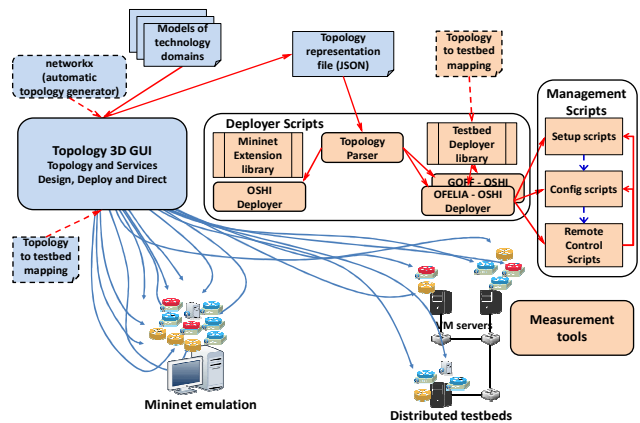

Figure 11. Mantoo enabled emulation workflow

The Topology3D exports the representation of the views (topology and services) in a JSON format, which becomes the input for the deployment phase. We have integrated the Networkx [27] tool which allows generating random data plane topologies with given characteristics.

B. *Deployment phase*

The deployment phase translates the designed topology into the set of commands that instantiate and configure the nodes and the services for a given experiment. This phase can target different execution environments for the experiments, by means of a specific "Deployer". So far, we targeted one emulator (Mininet) and four distributed SDN testbeds (the OFELIA testbed [10], the GÉANT OpenFlow Facility – GOFF [41], the GÉANT Testbeds Service – GTS [42] and a private testbed called Netgroup SDN Testbed – NeST [9]).

Technically, the deployment phase is performed by a set of python scripts (Topology Deployer) that parse the JSON file with the representation of the views and produce further scripts (mostly shell scripts). The proper execution of these scripts deploys the experiment either over Mininet or over a distributed SDN testbed. The Testbed Deployer and the Mininet Extensions are Python libraries that are used by the actual Deployers. The Mininet Extensions library is tailored for the Mininet emulator, while the Testbed Deployer currently supports the four above mentioned testbeds and it can be easily extended to support additional ones.

*1) Mininet Extensions*

By default, Mininet only provides the emulation of hosts and switches. We enriched Mininet introducing an extended host, capable of running as a router and managed to run the Quagga and OSPFD daemons on it. The extended host includes Open vSwitch, as needed to realize the OSHI node. Another enhancement to the default Mininet setup depends on our requirement to reach the emulated nodes via SSH from an external, "non-emulated" process. For this purpose, we introduce a fictitious node in the root namespace of the hosting machine that is connected to the emulated network and works as relay between the emulated world of Mininet and the "real" world of the hosting machine. The details on the specific Mininet deployment architecture can be found in [9]. The Mininet Extensions library is able to automate all the aspects of an experiment. This includes the automatic configuration of IP addresses and of dynamic routing (OSPF daemons) in all nodes, therefore relieving the experimenter from a significant configuration effort. As for the software design, the library extends Mininet providing new objects and API that seamlessly integrate with existing Mininet objects.

*2) Deployment over distributed SDN testbeds*

We implemented and tested a Deployer for each of the four distributed SDN testbeds listed above. The OFELIA and GOFF testbeds share a similar architecture as they are based on the OCF (OFELIA Control Framework) [10]. These two testbeds manage differently the out-of-band connectivity. Specifically, in the OFELIA testbed there is a management network with private IP addresses, while in the GOFF testbed all the virtual machines use a public IP address. The OFELIA testbed slice we used is hosted in the CREATE-NET island, composed by 8 OpenFlow capable switches and 3 Xen [44] Virtualization Servers for the experimental Virtual Machines (VMs). The GOFF testbed offers five sites, each one hosting two servers, which respectively run the OF equipment (based on OVS) and Xen, for hosting the VMs. The GOFF testbed supports all the OSHI services (IP VLLs, PW and VSS). In the OFELIA testbed the PW and VSS services cannot be deployed due to old Linux kernels which do not support network namespaces. The GTS testbed is distributed on a number of locations interconnected by the GÉANT core network [43]. It is managed by OpenStack, each site includes a KVM Virtualization Server and a physical OpenFlow capable switch. Finally, NeST is a small private testbed located at University of Rome Tor Vergata, composed by three servers, each one running both a KVM Virtualization Server and a switch based on OVS.

The Management Scripts automate and facilitate the setup, configuration and the deployment of an experiment. They relieve the experimenter from tedious and error prone activities. As shown in Figure 11, the Testbeds Deployer Scripts automatically produce the configuration files that are given in input to the Management Scripts for emulating a given topology, composed of access and core OSHI nodes (OSHI-PE and OSHI-CR) and end points (CEs and SDN controllers). This includes the automatic configuration of IP addresses and of dynamic routing daemons (OSPF) on all nodes, saving a significant time for the node configuration. Each node (CR, PE or CE) is mapped into a different VM running in a Virtualization Server of a given testbed. Two mechanisms can be used to map an emulated node on a VM: 1) a resource file (called "topology-to-testbed") with a list of IP addresses of available VMs can be given to the Deployer, which automatically choses the VMs for the emulated nodes; 2) it is possible to manually assign the target VM (identified by its IP address) for an emulated node, either editing a mapping file or graphically using the Topology3D GUI.

A management host coordinates the overall process, usually also executing the Deployer scripts. The management host and the VMs communicate over a management network. The configuration files generated by the Deployers scripts are uploaded on a repository reachable by the VMs (e.g. a webserver running on the management host). During the deployment process these files are downloaded by each VM belonging to the experiment.

The Management Scripts are logically decomposed in Remote Control Scripts, Setup Scripts and Config Scripts:

- The Remote Control Scripts, based on Distributed SHell (DSH), are used by the management host for distributing and executing remote scripts and commands. They enable root login without password, avoid initial ssh paring and configure the DSH in the management VM. Once DSH has been properly configured with the IP of the VMs belonging to the experiment, it can run commands on a single machine, on a subset, or on all the deployed VMs. It is also possible to execute parallel commands speeding up the deployment.
- The Setup Scripts turn a generic VM provided by the testbed into an emulated node (CR, PE, CE or controller), installing and configuring the needed software modules.
- The Config Scripts configure a specific experiment and its topology, setting up the link (tunnels) among the VMs.

In order to replicate an experimental topology emulating the network links among CRs, PEs and CEs an overlay of Ethernet over UDP tunnels is created among the VMs, as shown in Figure 12 for the OFELIA and GOFF testbeds. A target overlay topology is shown in the higher part of the figure, while the physical testbed is shown in the bottom part, in this example it is constituted by two Virtualization Servers connected by a set of OpenFlow switches. Each element of the overlay topology (node, host or SDN controller) is mapped on a different VM that can be run in one of the Virtualization Servers, as shown in the middle part of the figure. The red thick lines represent the UDP tunnels among the VMs that are setup in order to map the links of the overlay topology. The underlying connectivity among the VMs has to be managed by the Testbed SDN Controller. In case of GTS and NeST the deployment is simplified because the underlying connectivity among the VMs is automatically provided by the testbed management infrastructure.

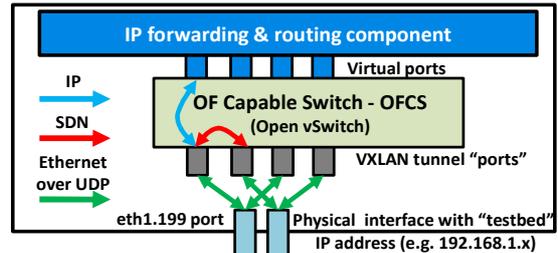

Figure 13. Implementing VXLAN tunnels using Open vSwitch (OVS)

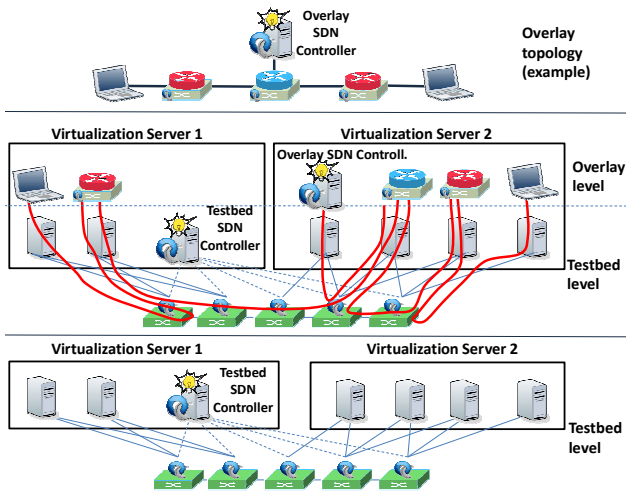

Figure 12. Deploying an overlay topology over the OFELIA/GOFF testbeds

A first option to build the tunnels is to use the user space OpenVPN tool (with no encryption). The performance is poor, as performing encapsulation in user space is very CPU intensive. A possible approach to enhance performance is to rely on specific hardware and/or on software modules on optimized I/O library like Intel DPDK [30]. We prefer a solution that is applicable on generic Linux devices, so we consider an approach based on the VXLAN tunnels [33] provided by Open vSwitch. OVS implements VXLAN tunnels in kernel space [32], dramatically improving performance with respect to OpenVPN. The design of the VXLAN tunneling solution for OSHI over a distributed testbed is reported in Figure 13. We only use VXLAN as a point-to-point tunneling mechanism (the VXLAN VNI identifies a single link between two nodes) and we do not need underlying IP multicast support, as in the full VXLAN model. The OF Capable OVS is also used to perform encapsulation and de-capsulation of VXLAN tunnels. Each tunnel corresponds to a port in the switch

C. *Control phase (running the experiments)*

In the Mininet based experiments it is possible to open consoles on the emulated nodes using the web GUI of the Topology3D. The consoles show the output generated by the ssh processes connected to the nodes (deployed in the Mininet emulator). The generated output is conveyed to the terminal shell running in the experimenter browser, leveraging the WebSocket API, where each terminal has a separate WebSocket channel. The same functionality for the experiments over the distributed testbeds is currently under development.

D. *Measurement Phase*

In order to automate as much as possible the process of running the experiments and collecting the performance data over distributed testbeds we have developed an object oriented multithreaded Python library called Measurement Tools. The library offers an intuitive API that allows the experimenter to "program" his/her tests. Using the library we can remotely (through SSH) run the traffic generators (iperf) and gather load information (CPU utilization) on all nodes (VMs). As for the load monitoring, taking CPU measurements from within the VMs (e.g. using the *top* tool) does not provide reliable measurements. The correct information about the resource usage of each single VM can be gathered from the virtualization environment, for example on Xen based systems we relied on the *xentop* tool, which must be run as root in the Xen based Virtualization Server. Therefore, for the OFELIA environment we have developed a python module that collects CPU load information for each VM of our interest in the Xen server using *xentop* and it formats it in a JSON text file. The Measurement Tools retrieve the JSON file from the python module with a simple message exchange on a TCP socket. In the GOFF environment the measurement data are provided through a Zabbix interface [46]. A python module gathers the data from the Zabbix API. In the KVM based NeST testbed, we relied on the *virt-top* tool.

The Measurement Tools provide a general framework that can be easily adapted to different needs. Currently we have developed tools able to generate UDP traffic and to gather CPU load information from the virtualization environment. An experimenter can easily extend this framework to run his/her tests and collect the measures of interest.

VII. PERFORMANCE EVALUATION ASPECTS

In this section we analyze some performance aspects of the OSHI prototype implementation over distributed SDN testbeds. The openness of the OSHI solution makes it possible to design and implement new services based on the SDN paradigm and run experiments to validate them and/or to compare different implementation options. Thanks to the Mantoo suite, an experimenter can deploy a large scale network over a distributed testbed. In our view the added value provided by OSHI/Mantoo will be the opportunity to get feedback on Control Plane design issue from the implementation and the experiments.

On the other hand in this section we focus on some Data Plane aspects of our prototype implementation. The rationale for this evaluation is to provide an indication on the scalability of the emulation approach in distributed testbeds made up of Linux Virtual Machines running on typical Virtualization

Servers. It is not our purpose to assess Data Plane forwarding performance for a production ready solution working at line speed in the core of ISPs' WANs. This type of evaluation will be needed if OSHI will be ported over the so called *white box switches*, high performance forwarding equipment with an open Operating System that can be customized by third-party developers, but this is for future work.

The first two experiments (sections VII.A, VII.B) have been performed over an OFELIA testbed. We used the iperf tool as traffic source/sink in the CE routers and generate UDP packet flows from 500 to 2500 packet/s. In these experiments the UDP packet size was 1000 bytes (using UDP packets ranging from 100 bytes to 1400 bytes, the performance has been only influenced by the packet rate). We evaluated the CPU load in the PE routers with our *xentop* based Measurement Tools. We executed periodic polling and gathered the CPU load of the monitored VMs. In each run we collected 20 CPU load samples with polling interval in the order of two seconds: the first 10 samples are discarded and the last 10 are averaged to get a single CPU load value. Then we evaluated the mean and the 95% confidence intervals (reported in the figures) over 20 such runs. The experiment in section VII.C has been executed on the NeST testbed, shown in Figure 16. The above described methodology has been used, but the generated packet rate ranged from 12.5 kp/s to 62.5 kp/s, with UDP packet size of 100 bytes, we evaluated CPU load both in PE and CR OSHI nodes, using the *virt-top* tool. Finally, the experiments in sections VII.D and VII.E have been performed on the GOFF testbed.

### A. *Best Effort IP performance in OSHI*

With reference to the architecture in Figure 2, we compared the forwarding performance of IP Best Effort packets in OSHI (where each packet crosses the Open vSwitch two times, marked as "OSHI IP" in Figure 14) with plain IP forwarding (the Open vSwitch is removed and the OSHI node interfaces are directly connected to IP forwarding engine, marked as "ROUTER IP"). In the next section, we refer to the OSHI-IP case as "No-Tunnel", as no tunneling mechanism is used. This experiment is not automatically deployed using the Topology3D and Deployer, and we setup a limited topology with two CE nodes and two OSHI nodes. In the experiment results (see [9] for details) we can appreciate a CPU load penalty for OSHI IP forwarding with respect to ROUTER IP forwarding ranging from 11% to 19% at different rates. The theoretical CPU saturation rate for plain ROUTER IP forwarding is in the order of 14000 p/s. OSHI IP forwarding reduces the theoretical CPU saturation rate to something in the order of 12500 p/s (corresponding to 11% performance penalty).

### B. *Performance comparison of tunneling mechanisms*

In this experiment we evaluated the processing overhead introduced by the tunneling mechanisms (OpenVPN and VXLAN) used to deploy the overlay experimental topologies over distributed SDN testbeds. We considered the same topology of the previous subsection.

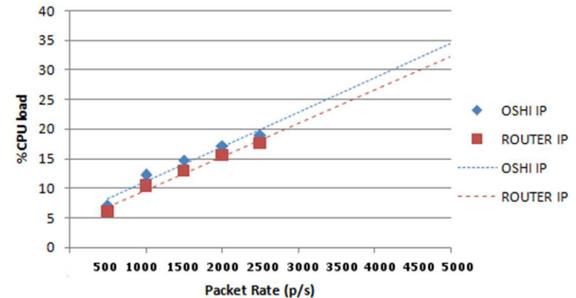
Figure 14. Best Effort IP forwarding performance.

Figure 15 compares the CPU load for OSHI IP forwarding in the OpenVPN, VXLAN and no tunneling scenarios. It can be appreciated that VXLAN tunneling adds a reasonably low processing overhead, while OpenVPN tunneling would dramatically reduce the forwarding capability of an OSHI node in the testbeds. The theoretical CPU saturation rate for OpenVPN tunneling is in the order of 3500 p/s, which is 4 times lower than in the no tunneling case. The theoretical CPU saturation rate for VXLAN tunneling is only ~8% lower than the no tunneling case, showing that VXLAN is an efficient mechanism to deploy overlay topologies.

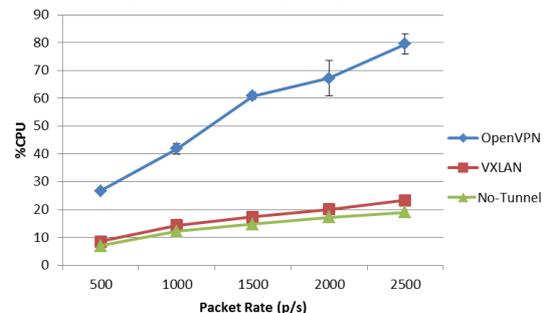
Figure 15. CPU Load for different tunneling mechanisms.

### C. *Performance comparison of different forwarding approaches over the distributed SDN testbed*

In this experiment we evaluated the processing load of different forwarding approaches over the distributed SDN testbeds considering the topology shown in Figure 17. For the OSHI solution, we considered IP forwarding (OSHI IP) and SBP forwarding (OSHI VLL). Then we assumed plain IP forwarding as a reference (ROUTER IP).

We executed the performance tests of OSHI IP, OSHI VLL and ROUTER IP using the VXLAN tunneling solution and collected the CPU load both for the access PE node and the first CR node (see results in Figure 18). In case of plain IP forwarding (ROUTER IP) the packets have to cross the Open vSwitch which handles the VXLAN tunneling (see Figure 13), therefore as expected there is no advantage with respect to OSHI IP. The OSHI VLL solution is the least CPU intensive as it exploits MPLS label switching in the Open vSwitch. The CPU performance penalty of OSHI IP forwarding w.r.t. OSHI VLL is less than 10%. The CPU loads for PE and CR are different in absolute values because the respective VMs are mapped in two different Virtualization Servers with different processors. In the experiment, a physical core of the Virtualization Servers was exclusively allocated to each VM. For the PE node the theoretical CPU saturation rate is in the order of 320 kp/s for OSHI VLL, while for the CR node hosted on the more performant server the theoretical CPU saturation rate is in the order of 1 Mp/s.

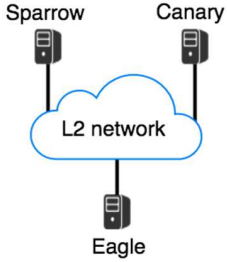
Figure 16. Physical network in the NeST testbed

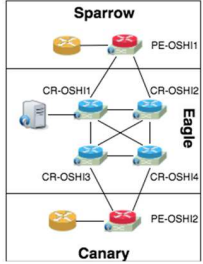
Figure 17. Overlay network for the experiment on NeST

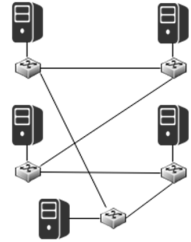
Figure 19. GOFF Physical network

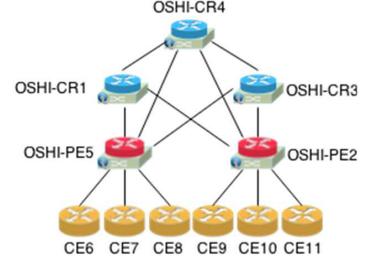
Figure 20. Overlay network for the experiment on GOFF

In the experiment results (see Figure 21) we can appreciate a CPU load penalty for OSHI PW forwarding with respect to OSHI VLL forwarding in the order of 15%-21%. Apparently, the CPU load penalty is decreasing in relative terms at higher CPU load. These results shows the potential improvements that could be achieved by natively supporting EoMPLS tunneling in the switches instead of using the developed ACE and the GRE encapsulation.

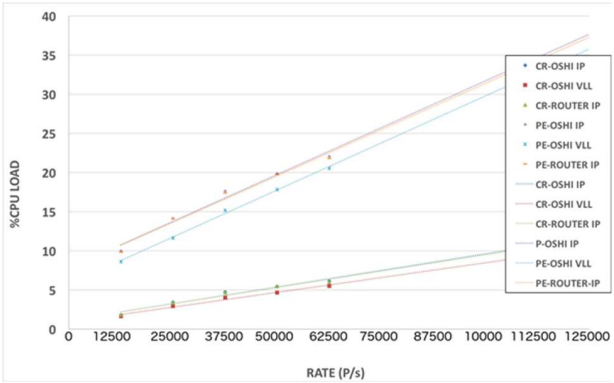
Figure 18. CPU load with VXLAN tunneling.

D. *Performance evaluation of encapsulation for PW service*

In this experiment we evaluated the performance penalty introduced by the encapsulation mechanism implemented for the PW service (section IV.D). We have performed this experiment over the GOFF testbed (physical topology is represented in Figure 19) using the overlay topology shown in Figure 20. As usual, the iperf tool has been used as traffic source/sink in the CE routers and generates UDP packet flows. We evaluated the CPU load in the OSHI-PE5, with a periodic polling approach. A sample is provided by Zabbix every minute, representing the average calculated in this period with 1-second-interval samples. For each load level (packet rate) we executed a single run of 7 minutes and collected 7 CPU load values, the first 2 are discarded and the last 5 are averaged to get a single CPU mean load value. Then we evaluated the relative standard deviation (RSD) to ascertain the reliability of the results. The RSD is always smaller that 5% in all runs.

In the PE nodes, the implementation of the IP VLL service is based on the design shown in Figure 2, while the PW service considers the architecture described in Figure 7. We wanted to estimate the overhead introduced by the ACE and by the operations of the GRE tunnel. We generated UDP packet flows with a rate ranging from 2000 to 18000 packet/s (datagram size is 1000 byte as usual). The core topology is represented in Figure 20. In the experiment, 3 CEs, acting as traffic sources/sinks, were connected to each PE. This was needed because the generation rate of a single CE in this specific testbed setup was at most 6000 packet/s, to keep the CPU load of the CE VMs under a safety threshold.

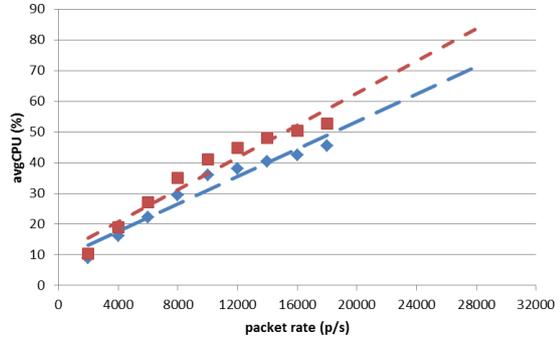
Figure 21. CPU load for different OSHI services.

E. *Performance analysis of OVS internal mechanisms.*

In this section, we shortly report about two experiments that concern the evaluation of OVS internal mechanisms. These experiments do not directly concern OSHI, but they support the choice of OVS as the software based OpenFlow capable switch integrated in OSHI node and show the effectiveness of the proposed Mantoo platform for the setup, deployment and control of the experiments and the collection of performance results. For space reasons, the detailed results have not been included and can be found in [47].

The first experiment investigates the impact of the kernel flow cache implemented in OVS. In the OVS architecture, the first packet of a flow arriving at a node is forwarded to a Linux user space process, while the following packets are using a flow cache in the kernel. OVS performance is optimal as long as the packets are forwarded using the kernel flow cache. For the same traffic pattern we measured 40% CPU utilization for kernel cache processing and 94% utilization for user space processing. For the OSHI solution, we gathered the design insight that the number of active SBPs should remain within the limit of the kernel flow table. We evaluated (details in [47]) how many flow table entries are needed for an IP VLL or L2 PW service, so that we relate the dimension of the flow table with the maximum number of service instances.

The second experiment evaluated how the number of active flows in the flow tables influences the forwarding performance of OVS. The comforting result is that increasing the number of active flows in the tables does not influence the forwarding performance. This is obviously valid as long as the active flows are less than the size of the tables. The results is a prove of the efficient implementation of flow lookup mechanisms, at least for the traffic patterns that we have used in our experiments.

VIII. RELATED WORK

Pure SDN solutions based on SDN capable switches interconnected with a centralized controller have been demonstrated both in data-centers and in geographically distributed research networks, such as OFELIA [10] in EU, GENI [11] and Internet2 [12][13] in US. To the best of our knowledge, these solutions do not integrate L3 routing within the SDN capable L2 switches. We argue that an ISP network requires a more sophisticated approach that can natively interwork with legacy IP routers and IP routing protocols. As stated in [7], a hybrid SDN model that combines SDN and traditional architectures may "sum their benefits while mitigating their respective challenges". Some recent works address the hybrid IP/SDN networking from different perspectives.

In [14] the authors presented an Open Source Label Switching Router that generates OSPF and LDP packets using Quagga. The node computes the MPLS labels that are then installed in the switches using the OpenFlow (OF) protocol. This architecture does not exploit a logically centralized controller. Instead, it considers a traditional distributed control plane, while it uses OF only locally in a node to synchronize the FIBs and to program the data plane.

RouteFlow [15] creates a simulated network made of virtual routers at the top of a SDN controller. The simulated network is a copy of the physical one. The controller uses the BGP protocol to interact with routers of neighbor domains and it simulates intra domain protocols (OSPF, IS-IS) between the virtual routers. A traditional IP routing engine (Quagga [16]) computes the routing tables that are eventually installed into the physical nodes via the OF protocol. The Cardigan project [18] is based on a fork of RouteFlow. Cardigan realized a distributed router based on RouteFlow concepts and deployed it in a public Internet exchange, showing the applicability of SDN/OpenFlow in a production context. The "SDN-IP" solution proposed in [19] follows similar principles. It is based on the ONOS SDN controller [20] and it also interacts with external domains using BGP. Differently from RouteFlow, the controller does not instantiate virtual routers to simulate the exchange of intra domain routing protocols, but it centralizes the routing logic for better efficiency.

Compared with these works, our solution assumes that the physical nodes still deal with basic IP routing, thus achieving resilience for basic IP connectivity based on standard IP routing and easier interoperability with non-OF devices in the core network. On top of the basic routing, the SDN/OpenFlow controller can instruct the hybrid IP/SDN nodes to perform SDN based forwarding for specific traffic flows. This idea of supporting such hybrid nodes is already included in the OpenFlow specifications since the first version of the protocol. Two types of devices are considered: OF-only and OF-hybrid which can support both OF processing and standard L2/L3 functionalities. Currently, only proprietary hardware switches implement the hybrid approach offering also L3 standard routing capabilities. OSHI represents a fully Open Source OF-hybrid solution designed to be flexible and scalable, so as to facilitate experimentation on hybrid IP/SDN networks at large scale.

The Google B4 WAN [21] is an integrated hybrid IP SDN solution, and it has likely been the first application of the SDN approach to a large-scale WAN scenario. In the B4 solution the traditional distributed routing protocols coexist with a SDN/OpenFlow approach. In particular, the B4 WAN sites are interconnected with traditional routing and the SDN-based centralized Traffic Engineering solution is deployed as an overlay on top of basic routing. Differently from the OSHI solution, the routing protocols are processed by servers external to the switches. Google B4 solution is proprietary and it is highly tailored to the needs of their specific scenario, composed of few large sites that needs to be interconnected. As such, it does not represent a typical ISP WAN network, made up by a large number of geographically distributed nodes. On the other hand, OSHI is designed as a generic and open solution for hybrid IP/SDN networks.

This work significantly extends the preliminary results described in [5]: 1) the implementation of SDN based paths is based on MPLS labels rather than VLAN tags, solving the scalability issues; 2) in addition to the IP VLL service the proposed solution offers the L2 PW service and the Virtual Switch Service on top of it; 3) the detailed design and implementation aspects of an OSHI node are described; 4) the Mantoo platform has been extended, for example it now supports remote consoles on the emulated Mininet nodes using the web GUI; 5) the experiments have been validated again with the new MPLS based implementation. A demo of the Mantoo platform has been presented in [48].

IX. CONCLUSIONS

In this paper we have presented a novel architecture and implementation of a hybrid IP/SDN (OSHI) node. The OSHI data plane supports the coexistence of best effort IP forwarding and SDN based forwarding using MPLS labels. The traditional distributed MPLS control plane is not needed anymore, as all MPLS circuits (Label Switched Paths, now termed *SDN Based Paths*) are established by means of the SDN controller. We have shown the implementation of IP VLL and Layer 2 Pseudo Wire (PW) services. On top of the L2 PW service we also have built a layer 2 Virtual Switch Service (VSS), closely resembling the layer 2 VPLS solution over MPLS. Using the SDN approach, all complex control plane functions that take decisions (e.g. optimal tree evaluation) and enforce that decisions (e.g. creation of PWs) are executed outside the OSHI network nodes. Results of performance tests executed both in single-host emulators (Mininet) and in distributed SDN testbeds have shown that OSHI is suitable for large-scale experimentation settings.

We have described Mantoo, a suite of supporting tools for experiments with OSHI based services. It includes an extensible web GUI framework for designing and validating a topology, called Topology3D. The topology is automatically deployed either on Mininet or on distributed testbeds. Execution and Measurement tools simplify running the experiments and collecting performance measurements.

Developed according to an Open Source model, the OSHI prototype and the Mantoo suite are valuable tools that enable further research and experimentation on novel services and architecture in the emerging hybrid IP/SDN networks.

So far, we presented our implementation of the OSHI architecture mostly as an experimenter tool. It allows to easily configure VMs as hybrid IP/SDN nodes and perform experiments at relatively large scales using Mininet emulator or resources over distributed testbeds. On the other hand, we recently started working on an implementation of the OSHI architecture on white box switches [4], in particular using the P-3922 10Gbe switch from Pica8. This work goes into the direction of implementing OSHI in devices that can perform switching and routing at line speed over production networks, closing the gap between SDN research and real world networks. Details on these white box switches experiment scenarios and results are available at [6].

## X. ACKNOWLEDGMENTS

This work was partly funded by the EU in the context of the projects: GÉANT GN4 Phase 1, SUPERFLUIDITY (5G PPP), DREAMER [35] (a beneficiary of the GÉANT Open Call research initiative of the GN3plus project).